\documentclass[fleqn,usenatbib]{mnras}

\usepackage{newtxtext,newtxmath,tablefootnote}

\usepackage[T1]{fontenc}
\usepackage{tabularx}

\DeclareRobustCommand{\VAN}[3]{#2}
\let\VANthebibliography\thebibliography
\def\thebibliography{\DeclareRobustCommand{\VAN}[3]{##3}\VANthebibliography}

\usepackage{graphicx}	
\usepackage{amsmath}	

\title[Constraining GW velocities]{Constraining gravitational wave velocities using gravitational and electromagnetic wave observations of white dwarf binaries}

\author[T. Y. Cao et al.]{
Tian-Yong Cao,$^{1,2}$
Ankit Kumar,$^{1}$\thanks{E-mail: kumarankit@ihep.ac.cn}
and Shu-Xu Yi$^{1}$\thanks{E-mail: sxyi@ihep.ac.cn}
\\
$^{1}$Key Laboratory of Particle Astrophysics, Institute of High Energy Physics, Chinese Academy of Sciences,\\ 19B Yuquan Road, Beijing 100049, People’s Republic of China\\
$^{2}$University of Chinese Academy of Sciences, Chinese Academy of Sciences,\\ Beijing 100049, People’s Republic of China
}

\date{Accepted 2024 July 16. Received 2024 June 24; in original form 2024 April 8}

\pubyear{2024}

\begin{document}
\label{firstpage}
\pagerange{\pageref{firstpage}--\pageref{lastpage}}
\maketitle

\begin{abstract}
\noindent Although the general theory of relativity (GR) predicts that gravitational waves (GWs) have exactly the same propagation velocity as electromagnetic (EM) waves, many theories of gravity beyond GR expect otherwise. Accurate measurement of the difference in their propagation speed, or a tight constraint on it, could be crucial to validate or put limits on theories beyond GR. The proposed future space-borne GW detectors are poised to detect a substantial number of Galactic white dwarf binaries (GWDBs), which emit the GW as semi-monochromatic signals. Concurrently, these GWDBs can also be identified as optical variable sources. Here we proposed that allocating a GWDB's optical light curve and contemporaneous GW signal can be used to trace the difference between the velocity of GW and EM waves. Simulating GW and EM wave data from 14 verification binaries (VBs), our method constrains propagation-originated phase differences, limiting the discrepancy between the speed of light ($c$) and GW ($c_{GW}$). Through the utilization of LISA's design sensitivity and the current precision in optical observation on GWDB, our study reveals that a four-year observation of the 14 recognized VBs results in a joint constraint that confines $\Delta c/c$ ($\Delta c = c_{\mathrm{GW}} - c$) to the range of $-2.1\times10^{-12}$ and $4.8\times10^{-12}$. Additionally, by incorporating this constraint on $c_{\mathrm{GW}}$, we are able to establish boundaries for the mass of the graviton, limiting it to $m_{\mathrm{g}}\le3\times10^{-23}\,e\mathrm{V}/c^{2}$, and for the parameter associated with local Lorentz violation, $\bar{s}_{00}$, constrained within the range of $-3.4\times10^{-11}\le\bar{s}_{00}\le1.5\times10^{-11}$.
\end{abstract}




\section{Introduction}
	
	Following Einstein's proposal of the general theory of relativity (GR) in 1915, he theoretically forecasted the existence of gravitational waves (GW) \citep{a,b}. According to his prediction, the speed of GW was expected to be precisely equal to the speed of light \citep{c}. Further exploration into the realm of gravity has prompted certain researchers to propose alternative theories, a few of which involve modifications to the speed of GW. For example, if the graviton possesses a non-vanishing mass, its velocity would be dispersed in a vacuum as \citep{d}:
       
	   \begin{equation}
            \frac{c_{\mathrm{GW}}}{c}=(1-\frac{m_{\mathrm{g}}^{2}c^{4}}{E^{2}})^{1/2}\approx1- 
            \frac{1}{2}\frac{m_{\mathrm{g}}^{2}c^{4}}{E^{2}},
		\label{v-f}
	   \end{equation}
 
        \noindent where $m_{g}$ is the mass of graviton; $c$ and $c_{\mathrm{GW}}$ are the speed of light and GW respectively; and $E=hf$ is the energy of the graviton, where $h$ is the Planck constant and $f$ is the frequency of GW. The predicted dispersion pattern of gravitational waves (GW) has served as a means to investigate the mass of gravitons, causing phase differences among diverse frequency components of GW, and thereby causing alterations in its waveform , formally a 1 post-Newtonian order term \citep{e,f}:
        
	   \begin{equation}
	    \Phi(f)=-\frac{\pi D\,c}{\lambda_{\mathrm{GW}}^{2}(1+z)\,f},
	   \end{equation}
 
        \noindent where $\lambda_{\mathrm{GW}}=h/(m_{\mathrm{g}}c)$ is the graviton's Compton wavelength, $ z $ is the cosmological redshift and $D$ is the cosmological distance as defined explicitly in \cite{e}. By aligning the observed waveform with its theoretical projection \citep{r}, we can establish confidence intervals for the graviton's wavelength or mass. Employing a similar methodology, the LIGO and Virgo collaboration utilized gravitational wave events GW150914, GW170104, and GW170817 to establish upper limits for the mass of gravity, which were set at $ 1.2\times10^{-22}\,e\mathrm{V}/c^{2} $, $ 7.7\times10^{-23}\,e\mathrm{V}/c^{2} $, $ 9.51\times10^{-22}\,e\mathrm{V}/c^{2} $ \citep{f,g,h}, respectively. Correspondingly, at 1 kHz, the upper limits on $\Delta c/c$ from the above mentioned GW events stands at $<4.2\times10^{-22}$, $1.7\times10^{-22}$, and $2.64\times10^{-22}$, where $ \Delta c=c_{\mathrm{GW}}-c $. This constraint is notably stringent, nevertheless presuming $c_{\mathrm{GW},\infty}=c$, wherein $c_{\mathrm{GW},\infty}$ represents the velocity of gravitational waves at an infinitely high frequency. Subsequent discussion will unveil that additional methods, independent of this particular assumption, are still required. Moreover, if alternative methods can impose constraints on $c_{\mathrm{GW}}$ in a much lower frequency regime, it could, in turn, restrict the possible range of the graviton's mass.

        \begin{table}
            \caption{\textbf{A Summary outlining the constraints obtained by utilizing various methods.} Refs.: [1] \citep{f}, [2] \citep{g}, [3] \citep{h}, [4] \citep{m}, [5] \citep{s}, [6] \citep{k}, [7] \citep{wc}, [8] \citep{wd}. \\}
            \renewcommand{\tabcolsep}{0.35cm}
            \renewcommand{\arraystretch}{1.5}
            \centering
            \begin{tabular}{ccc}
            \hline
            Method & Result & Refs. \\ \hline
            GW dispersion & $\frac{\Delta c}{c}<1.7\times10^{-22}$ & 1,2,3 \\ 
            \hline
            \begin{tabular}[c]{@{}c@{}}time-lag:\\ GRB \& GW \end{tabular}  & $-3\times10^{-15}\le\frac{\Delta c}{c}\le+7\times10^{-16}$ & 4  \\ 
            \hline
            \begin{tabular}[c]{@{}c@{}}time-lag:\\ two GW detectors\end{tabular} &    $-0.03\le\frac{\Delta c}{c}\le+0.02$ & 5,6 \\ 
            \hline
            GWDBs \tablefootnote{It is an estimation of the proposed method, instead of constraints from real observation.} & $\frac{\Delta c}{c}<3\times10^{-12}$ & 7,8 \\ 
            \hline
            \end{tabular}
            \label{summary}
        \end{table}

	
	A theoretical framework known as the Standard Model extension (SME) characterizes general violations of Lorentz and CPT invariance in both general relativity and the standard model at attainable energies \citep{i,j}. In the context of SME, the GW velocity will be corrected to:
        
	\begin{equation}
		\frac{c_{\mathrm{GW},\pm}}{c}=1-\varsigma^{0}\pm|\vec{\varsigma}|.
	\end{equation}
 
        \noindent where $c_{\mathrm{GW},\pm}$ represents the velocity of the two polarizations of GW. If $|\vec{\varsigma}|$ is non-zero, birefringence will occur. With $\varsigma^{0}$ exerting notable dominance over the $|\vec{\varsigma}|$ term, as highlighted by \cite{i}, one can safely opt to exclude consideration of $|\vec{\varsigma}|$ in their calculationss \citep{k}. Consequently, our attention is solely on the deviation value $ \varsigma^{0} $ in this context. This approximation allows us to expand the GW velocity employing the spherical harmonic function: 
        
	\begin{equation}
		\begin{aligned}
                \frac{c_{\mathrm{GW}}}{c}(\alpha,\delta) &= 1+\frac{1}{2}\sum_{jm}^{}(-1)^{j}Y_{jm}(\alpha,\delta)\bar{s}_{jm} \\
			&=1+\sum_{j}^{}(-1)^{j}\big(\frac{1}{2}\bar{s}_{j0}Y^{j0}\\
			& +\sum_{m>0}{}[\Re\bar{s}_{jm}\Re Y_{jm}-\Im\bar{s}_{jm}\Im Y_{jm}]\big),
		\end{aligned}
		\label{SME}
	\end{equation}
 
        \noindent where $\Re x$ and $\Im x$ represent the real and imaginary parts of $x$. $Y_{jm}$ is a spherical harmonic function, therefore $m$ should be an integer satisfying $-j\le m\le j$. Within the SME framework, it is posited that $j\leq2$. In order to constrain the nine coefficients present in the above equation, a minimum of nine GW sources, each with their respective $c_{\rm GW}$ limits and corresponding sky positions, are needed. 
	
        There are several other ways to put constraints on the $c_{\mathrm{GW}}$. One of the stringent constraints arises from the time difference between GW170817 and GRB170817A \citep{l, la}. On 2017 August 17, the GW event GW170817 was observed by LIGO, and the associated gamma-ray burst(GRB) GRB170817A was observed independently by the Fermi Gamma-ray Burst Monitor(GBM). There is a $ 1.74\pm0.05 $s delay between the prompt emission of GRB170817A and the chirp time of GW170817 \citep{m}. The observed time lag between the GW170817's peak and the first photon of GRB 170817A was used to constrain on the fractional speed difference to $-3\times10^{-15}\le\Delta c/c\le+7\times10^{-16} $ \citep{n}. The result depends on the assumption of the intrinsic emission time difference between the GW and GRB, which is a debating presumption. \citep{o,p,q} A previous simulation study \citep{bb} showed that the joint detection of GW and its GRB prompt-emission counterpart will be a rare event. Therefore, we would expect obtaining a sample of $c_{\rm GW}$ from at least nine different sources with this method, to be challenging in a short time scale. 
	
        There are also other model/source-independent methods, such as using the time difference between two detectors at large distances to the same GW signal. For example, the LIGO at the Hanford and Livingston sites have a time difference of about 10 milliseconds. \cite{s} used a Bayesian approach, combining data from the first three LIGO GW observations, to constrain $c_{\mathrm{GW}}$ to $ -0.45\le\Delta c/c\le+0.42 $. Subsequently, \cite{k} further narrowed the interval to $ -0.03\le\Delta c/c\le+0.02 $ with more data and constrained the coefficient $\bar{s}_{00}$ within $ -0.2<\bar{s}_{00}<0.07$.

        \begin{table*} 
            \caption{\textbf{The observed parameters of VBs}: The contents in each column of the table are, in order, the name of the source, right ascension, declination, orbital period, GWDB masses, distance, orbital inclination, 4-year SNR in LISA, and references. Refs.: [1] \citep{y}, [2] \citep{w}, [3] \citep{aa}, [4] \citep{ab}, [5] \citep{ac}, [6] \citep{ad}, [7] \citep{ah}, [8] \citep{ai}, [9] \citep{aj}, [10] \citep{ak}.\\}
		\centering
            \renewcommand{\tabcolsep}{0.48cm}
            \renewcommand{\arraystretch}{1.5}
            \begin{tabular}{cccccccccc}
			\hline
			Source & $\alpha$ & $\delta$ & P & $m_{1}$ & $m_{2}$ & d & $i$ & SNR & Refs. \\
            & /deg & /deg & /s &  /$ M_{\odot}$ &  /$M_{\odot}$ & /kpc & /deg & (4 years) &  \\\hline
			J1539+5027  & 234.88 & 50.46  & 414.79 & 0.61   & 0.21   & 2.34   & 84.15                     & 143                         &   1   \\
			J0651+2844 & 102.89 & 28.74  & 765.5       & 0.26   & 0.5    & 0.933  & 86.9                      & 90.1                        &   2,3,4   \\
			J0935+4411 & 143.78 & 44.19  & 1188        & 0.312  & 0.75   & 0.645  & 60 & 44.9                        &   2,5   \\
			J0923-1218               & 140.96 & -12.31 & 252.36     & 0.344  & 0.19   & 0.262  & 60 & 31                          &   6   \\
			J1638+3500               & 249.61  & 35.00  & 1186.12    & 0.698  & 0.45   & 0.103  & 60 & 30                          &   6   \\
			J0130+5321               & 22.74 & 53.36  & 518.53     & 0.191  & 0.4    & 0.085  & 60 & 24                          &    6  \\
			J1738+2927               & 264.65 & 29.46  & 242.43     & 0.261  & 0.55   & 0.78   & 60 & 24                          &   6   \\
			J1115+0246               & 168.86 & 2.77  & 1071.79    & 0.446  & 0.26   & 0.899  & 60 & 16                          &   6   \\
			J0533+0209  & 83.38 & 2.15  & 1233.97  & 0.652  & 0.167  & 1.5    & 72.8                      & 15.2 &   7   \\
			J1401-0817               & 210.33 & -8.29 & 278.92     & 0.216  & 0.79   & 0.555  & 60 & 13                          &    6  \\
			J1048-0000               & 162.11  & -0.0158 & 521.12     & 0.169  & 0.62   & 0.707  & 60 & 6.3                         &   6   \\
			J0923+3028 & 140.94   & 30.47  & 3383.68     & 0.275  & 0.76   & 0.299  & 60 & 5.6  &    2,8  \\
			CD-3011223               & 212.82 & -30.88 & 4231.8      & 0.54   & 0.79   & 0.337  & 82.9                      & 4.9                         &   2,9   \\
			J1630+4233 & 247.63 & 42.555  & 2389.8      & 0.298  & 0.76   & 1.019  & 60 & 4.6                         &   2,10   \\ \hline
		\end{tabular}
		\label{base data}
	\end{table*}
	
        Another model-independent method yielding more stringent constraints involves utilizing GW and EM signals emitted by Galactic white dwarf binaries (GWDB) \citep{wc}. It's believed there are roughly 11.5 billion white dwarfs (WD) in our Galaxy \citep{t}, and a significant proportion of these will come together to form binary systems that steadily emit GW. In general, the frequency of these emitted GWs are below $10^{-3}$ Hz \citep{u,v} and cannot be detected effectively by ground-based GW detectors like LIGO, Virgo, and KAGRA. However, within a certain range, there are tens of millions of GWDBs that could be detected by space-borne GW detectors like LISA \citep{w}, TianQin \citep{wa, wb} or Taiji \citep{wba}. There are several known GWDBs, whose binary parameters are well measured with optical observation, and they can be effectively detected by space-borne GW detectors. Those GWDBs are usually referred to as verification GWDBs or verification binaries (VBs). 

	
	In 2000, \cite{wc} suggested comparing the potential phase difference between the GW and optical signals of GWDB to constrain the graviton's mass. In their work, they used the ratio between the sampling time and the total integration time to estimate the error in phase measurements. In addition, they focused on the errors caused by refraction due to interstellar medium refraction in the EM wave propagation processes and the initial phase difference at the source between the GW and EM signals. Later, \cite{wd} specifically discussed the phase error in GW measurements on GWDB, correcting the error to:
        
	\begin{equation}
            \Delta\phi_{\mathrm{GW}}=\frac{\alpha}{2}(\frac{\mathrm{S}}{\mathrm{N}})^{-1}[1+\mathcal{O}(\mathrm{S}/\mathrm{N})^{-1}],
	\end{equation}
 
        \noindent where $\alpha$ is a correction term from the initial phase difference from the GW and EM signals. Moreover, they estimated the constrain to the graviton's Compton wavelength as $\lambda_{\mathrm{g}}>35\times10^{12}\,\mathrm{km}$. They used analytical error propagation estimation and did not include the intrinsic phase discrepancy between GW and EM counterpart signals. All methods for measuring $\Delta c/c$ and the corresponding results (including the limits from real observation and the estimated level of limits from proposed methods) are summarized in Table \ref{summary}.
	
        Here in the present work, we use simulation to consider the phase error of EM and GW altogether with more sources and also include the intrinsic phase difference between EM and GW due to polarization angle. We simulate the signals of GW and EM emitted by the VBs, and then fit the phase difference between the received signals. we exclude the initial phase differences and obtain an estimation of the phase differences due to propagation, and thus constraints on $\Delta c/c$. The VBs utilized in this paper, along with their binary parameters, are presented in Table \ref{base data}.

        The adopted method will be described in detail in the next section. In Section 3, we will focus on the intrinsic phase difference between GW and EM. In Section 4, we will discuss the method of simulating GW data. The process of optical light curve signal simulation will be outlined in Section 5. Section 6 will demonstrate how we match the phases using a joint fit and unveil our findings. Sections 7 and 8 are dedicated to exploring the boundaries of physical quantities and discussing the improvements.
	
    \section{Methodology overview}

        \begin{table*}
            \caption{\textbf{The table of polarization phases and the main parameters with their uncertainties}: The data in each column of the table are, in order, the value and uncertainty of right ascension, declination, orbital inclination, and the polarization phases. The uncertainties of RA and DEC are from GAIA \citep{an, ao}, and the RA, DEC, orbital inclination and the uncertainties of inclination is from \citep{w}. The uncertainty of right ascension and declination is presented in the unit "mas" which refers to milliarcseconds.\\}
		\centering
            \renewcommand{\tabcolsep}{0.58cm}
            \renewcommand{\arraystretch}{1.5}
		\begin{tabular}{ccccccccc}
            \hline
			Source & $\alpha$ & $\Delta\alpha$ & $\delta$ & $\Delta\delta$ & $i$ & $\Delta i$ & $ \varphi_{0} $ & $ \Delta\varphi_{0} $ \\ 
                &  /deg & /mas & /deg & /mas & /deg & /deg & /rad & /rad \\\hline
			J1539+5027 & 234.88 & 0.4911 & 50.46  & 0.4823 & 84.15 & 0.64 & -1.1313 & 0.0422 \\
			J0651+2844 & 102.89 & 0.3093 & 28.74  & 0.2825 & 86.9  & 1    & -1.4257 & 0.0446 \\
			J0935+4411 & 143.78 & 0.4814 & 44.19  & 0.4117 & 60    & 2    & -0.8686  & 0.0178 \\
			J0923-1218 & 140.96 & 0.0446 & -12.31 & 0.0400 & 60    & 2    & -0.4168 & 0.0141 \\
			J1638+3500 & 249.61 & 0.0213 & 35.00  & 0.0281 & 60    & 2    & -0.2981 & 0.0111 \\
			J0130+5321 & 22.74  & 0.0210 & 53.36  & 0.0219 & 60    & 2    & -0.1993 & 0.0087 \\
			J1738+2927 & 264.65 & 0.1801 & 29.46  & 0.2340 & 60    & 2    & -0.6239 & 0.0174 \\
			J1115+0246 & 168.86 & 0.2711 & 2.77   & 0.2880 & 60    & 2    & -0.5295 & 0.0158 \\
			J0533+0209 & 83.38  & 0.2169 & 2.15   & 0.1978 & 72.8  & 1.4  & -0.8166 & 0.0334 \\
			J1401-0817 & 210.33 & 0.0679 & -8.29  & 0.0502 & 60    & 2    & 1.3545 & 0.0082 \\
			J1048-0000 & 162.11 & 0.1692 & -0.016 & 0.1278 & 60    & 2    & -0.3598 & 0.0128 \\
			J0923+3028 & 140.94 & 0.0471 & 30.47  & 0.0370 & 60    & 2    & 1.5140 & 0.0038 \\
			CD-3011223 & 212.82 & 0.0442 & -30.88 & 0.0457 & 82.9  & 0.4  & -1.2674 & 0.0157 \\
			J1630+4233 & 247.63 & 0.1615 & 42.55  & 0.1768 & 60    & 2    & -0.3158 & 0.0115 \\ \hline
		\end{tabular}
		\label{pol phase}
	\end{table*}
	
        The time variation of the strain of the two polarization modes of GW for a non-evolving circular GWDB can be approximated by the quadrupole moment, which is derived by \cite{c} to vary sinusoidally. The response in the detector is a linear combination of the two modes in a certain proportion. \citep{wa} This proportion, known as the antenna pattern, will be provided with its specific expression in the next section. Therefore, the response in the GW detector can be represented with:
        
        \begin{equation}
		h(t)\propto\sin(2\pi f_{\rm{GW}}t+\phi_{0,\rm{GW}}),
        \end{equation}
 
        \noindent while the variance in the EM properties can be represented with:
        
        \begin{equation}
		A(t)\propto\sin(2\pi f_{\rm{EM}}t+\phi_{0,\rm{EM}}),
        \end{equation}
 
        \noindent with $f_{\rm{GW}}=nf_{\rm{EM}}$. For the WDs utilized in this paper, the EM properties that can be readily measured and exhibit a simple sinusoidal period include apparent magnitude \footnote{The a.m. is defined by $ m_{x}=-2.5\log_{10}\big(\frac{F_{x}}{F_{x,0}}\big) $, where $ F_{x} $ is the observed flux using spectral filter $x$ and $F_{x,0}$ is the reference flux (zero-point) for that photometric filter} (a.m.) and radial velocity \footnote{The r.v. is velocity projection in the radial direction. It is derived from the Doppler red/blue-shift in high quality spectroscopic observations, which need more observation resources and thus there are less available data points in Fig.3.} (r.v.). \citep{y,w,aa,ab,ac,ad,ah,ai,aj,ak}  It is worth noting that both the GW and a.m. data vary at twice the orbital frequency, while the r.v. varies at an equal frequency. Thus in the above equation, $n=1$ stands for a.m. and $n=2$ for r.v.. For the present scenario, we are considering the GW and EM signals received at the solar barycentre, assuming that the time and phase shift due to the relative motion between the detectors and the solar barycentre have been perfectly corrected \footnote{The relative motion between the detectors (optical telescopes and LISA) and the solar barycentre will introduce annual sinosoidal structures into the signals (due to geometrical delay and Doppler effect). If such noises are not corrected, any phase difference between GW and optical light curve can be dominated by such noise, and can not be used for our purpose. Luckily, such effects can be corrected as a standard procedure as the ephemeris of the Earth, the geolocation of the telescopes, and the orbital parameters of the LISA spacecrafts are known with high accuracies \citep{bh}. We, therefore, assume we can directly work with data with such effects corrected. }. The phase difference is:
        
        \begin{equation}
		\phi_{0,\rm{GW}}-n\phi_{0,\rm{EM}}=\Delta\phi_{\rm{int}}+\Delta\phi_{\rm{prop}},
        \end{equation}
 
        \noindent with $\Delta\phi_{\rm{int}}$ representing the intrinsic phase difference between EM wave and GW. In the present work, we set the phase of the EM wave as fiducial, and the main intrinsic phase offset is due to the polarization of the GW, which will be discussed in detail in the subsequent section. $\Delta\phi_{\rm{prop}}$ denotes the phase difference induced during the propagation, which reflects the potential difference between $c$ and $c_{\rm{GW}}$:
        
        \begin{equation}
            \Delta\phi_{\rm{prop}}=2D\frac{2\pi}{P}(1/c_{\rm{GW}}-1/c)\approx-\frac{4\pi D}{P}\frac{\Delta c}{c^2},
        \end{equation}

        \noindent where $P$ is the orbital period of the GWDB. Consequently, we can obtain the desired quantity i.e. $\Delta c$ as:
 
        \begin{equation}
            \Delta c=\frac{c^2P}{4\pi D}(n\phi_{\rm{EM}}-\phi_{0,\rm{GW}}+\Delta\phi_{\rm{int}}),
            \label{final}
        \end{equation}
 
        The following sections will illustrate our approach to restricting the phases separately, allowing us to then establish a restriction on $\Delta c$.
	
	\section{Intrinsic phase difference}
	
        The intrinsic phase difference $\Delta\phi_{\rm{int}}$ comprises several components, with the polarization phase being the most significant one. The GW strain recorded by the detector can be described as a linear combination of the two GW polarizations modulated by the detector’s response:
	
        \begin{equation}
            \begin{aligned}
            h(t)&=\frac{\sqrt{3}}{2}A_{+}F^{+}\cos 2\pi ft+\frac{\sqrt{3}}{2}A_{\times}F^{\times}\sin 2\pi ft\\
			&=\mathcal{A}\sin(2\pi ft+\varphi_{0}),
            \end{aligned}
        \end{equation}
 
        \noindent where $\mathcal{A}=\frac{\sqrt{3}}{2}\sqrt{(A_{+}F^{+})^{2}+(A_{\times}F^{\times})^{2}}$ is the combined amplitude, $\varphi_{0}=\arctan(-\frac{A_{+}F^{+}}{A_{\times}F^{\times}})$ is the polarization phase and $A_{+,\times}$ are the amplitudes of the two GW polarization, which can be represented as: 
        
        \begin{equation}
            \begin{aligned}
			&A_{+}(i)=\mathcal{A}_{0}(1+\cos^{2}i);\\
			&A_{\times}(i)=2\mathcal{A}_{0}\cos i,
            \end{aligned}
	\end{equation}

        where $i$ refers to the orbital inclination and $\mathcal{A}_{0}=\frac{2(G\mathcal{M})^{3/5}}{c^{4}d}(\pi f_{\rm{GW}})^{2/3}$ \citep{wa}. Here $\mathcal{M}$ is the chip mass and $d$ is the distance. The $F^{+,\times}$ are defined as the antenna pattern functions and can be mathematically represented as:
        
        \begin{equation}
            \begin{aligned}
			F^{+}(\theta_{s},\phi_{s},\psi_{s})&=\frac{1}{2}(1+\cos^{2}\theta_{s})\cos 2\phi_{s}\cos2\psi_{s}\\
			&-\cos\theta_{s}\sin2\phi_{s}\sin2\psi_{s};\\
			F^{\times}(\theta_{s},\phi_{s},\psi_{s})&=\frac{1}{2}(1+\cos^{2}\theta_{s})\cos 2\phi_{s}\sin2\psi_{s}\\
			&+\cos\theta_{s}\sin2\phi_{s}\cos2\psi_{s},
            \end{aligned}
	\end{equation}
 
        \noindent where $\psi_{s}$ is the polarization angle; $(\theta_{s},\phi_{s})$ are the latitude and longitude of the source in the detector’s coordinate frame, which can be transformed from Right Ascension (RA) and Declination (DEC) as in \cite{wa}. Further, we also estimated the uncertainties of the polarization phase $\varphi_0$ due to uncertainties in sky position, inclination, and polarization angle.
	
        We apply the Monte Carlo algorithm to calculate the phase uncertainty. We input randomly generated RA, DEC, inclination, and polarization angle values from 1000 normal distributions into the above equations. This process calculates the phase, allowing us to derive the polarization phase values with distribution. The main parameters with their uncertainties are shown in Table \ref{pol phase}, where the uncertainties of RA and DEC are from GAIA \citep{an, ao}, and other data is borrowed from \cite{w}. Besides, we take polarization angle $\psi_{s}=30^{\circ}$ with the uncertainty $\Delta \psi_{s}=10^{-3}$, which is supported from \cite{wa}. The polarization phase and its associated uncertainties are also presented in Table \ref{pol phase}. 

        \begin{table*} 
            \caption{\textbf{Optical data table of the VBs}: The data in each column of the table are, in order, the data volume, uncertainty, and amplitude of r.v. and a.m., $R$ of r.v. and a.m. Refs.: [1] \citep{y}, [2] \citep{w}, [3] \citep{aa}, [4] \citep{ab}, [5] \citep{ac}, [6] \citep{ad}, [7] \citep{ah}, [8] \citep{ai}, [9] \citep{aj}, [10] \citep{ak}.\\}
            \renewcommand{\tabcolsep}{0.49cm}
            \renewcommand{\arraystretch}{1.5}
            \centering
            \begin{tabular}{cccccccccc}
            \hline
            Source & $ N_{\mathrm{rv}} $ & $ \Delta A_{\mathrm{rv}} $ & $ A_{\mathrm{rm}} $ & $ N_{\mathrm{am}} $ & $ \Delta A_{\mathrm{am}} $ & $ A_{\mathrm{am}} $ & $ R_{rv} $ & $ R_{am} $& Refs. \\
            &  & /$ \mathrm{km}\cdot\mathrm{s}^{-1} $ & /$ \mathrm{km}\cdot\mathrm{s}^{-1} $ &  &  /mag &  /mag & $ \times10^{-2} $ & $ \times10^{-2} $ & \\
            \hline
			J1539+5027 & 12 & 43.10 & 550 &  &  &  & 2.26 &  & 1 \\
			J0651+2844 & 77 & 31 & 616.9  &  &  &  & 0.573 &  & 2,3,4 \\
			J0935+4411 & 55 & 28.73 & 198.5 &  &  &  & 1.95 &  & 2,5 \\
			J0923-1218 & 25 & 33.33 & 117 &  &  &  & 5.69 &  & 6 \\
			J1638+3500 & 25 & 25 & 89.5  &  &  &  & 5.59 &  & 6 \\
			J0130+5321 & 23 & 25 & 209.1 &  &  &  & 2.49 &  & 6 \\
			J1738+2927 & 16 & 20.83 & 372.7 &  &  &  & 1.40 &  & 6 \\
			J1115+0246 & 10 & 25 & 139.9 &  &  &  & 5.65 &  & 6 \\
			J0533+0209 & 35 & 100 & 618.7 & 5000 & 0.0275 & 0.134 & 2.73 & 0.290 & 7 \\
			J1401-0817 & 24 & 8.33 & 346.2  &  &  &  & 0.491 &  & 6 \\
			J1048-0000 & 17 & 16.67 & 312.8 &  &  &  & 1.29 &  & 6 \\
			J0923+3028 & 44 & 22 & 296 &  &  &  & 1.12 &  & 2,8 \\
			CD-3011223 & 95 & 11.67 & 376.6 &  &  &  & 0.318 &  & 2,9 \\
			J1630+4233 & 21 & 29.27 & 288.1 &  &  &  & 2.22 &  & 2,10 \\ \hline
            \end{tabular}
            \label{optical data}
        \end{table*}


        Additional intrinsic phase difference will be introduced between the GW and the a.m. signals, when the VB are not in tidal synchronization \citep{wc,wd}. In this case, the elongated axes of WDs can be misaligned with the line of masses. According to \cite{bc,bd,be}, detached VBs with orbital period less than 20 min can be considered tidally synchronized, where this additional intrinsic phase difference disappears. Therefore, when considering the a.m. light curves, we limit the sample to those detached VBs with period less than 20 min. It results in only one VB in the sample: J0533+0209. We like to note that, although there are theoretical arguments that this system can be in tidal synchronization, there lacks support from observation. We therefore present two version of results, one is with the assumption that J0533+0209 is in tidal synchronization and its a.m. light curve are used, the other one is when we do not make such assumption, and we use its r.v. data instead. 
	
	\section{GW simulation}
	
	\begin{figure} 
            \centering
		\includegraphics[width=\linewidth]{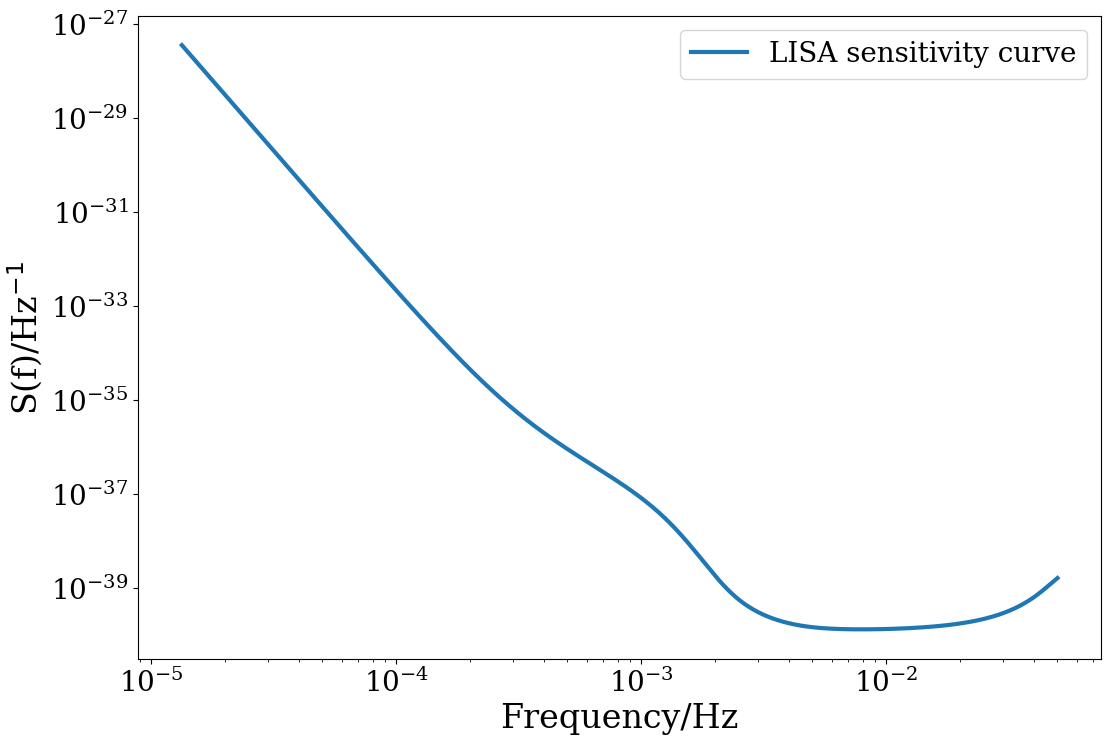}
            \caption{\textbf{The sensitivity curve of LISA depicted in \citep{am}.}}
		\label{sen curve}
	\end{figure}
	
        In the following section, we will employ simulation techniques to explore the accuracy with which we can determine the phase of gravitational waves. Here, we aim to implement several simplifications:
	
        First, we assume that the eccentricity of VBs has a negligible effect on the GW phase estimation. Typically, the trajectory of a GWDB deviates from a perfectly circular orbit and instead follows an elliptical trajectory characterized by a specific eccentricity. In practical terms, determining the eccentricity involves considerable uncertainty in measurement. However, the VBs utilized in this context are estimated to possess small eccentricity, allowing a circular orbit to serve as a reliable approximation. In the next section, we will observe that the radial velocities of all sources vary sinusoidally, indicating that their orbits are nearly circular.  Moreover, the gravitational waves (GW) emitted by elliptical orbits have evolved from their initial simple, single-frequency sine waveforms into a more complex composition, now characterized by a superposition of multiple harmonics, as discussed in \citep{c}, thereby intensifying the intricacy of the model. This paper, however, focuses on a more simplified scenario involving circular orbits. Our aim here is to test the fundamental principles of this novel method.  Integrating the effects of eccentricity into the application of this method while fitting the data with the model remains a straightforward task. Thus, considering the aim of model simplification, we've opted to streamline these gravitational wave database sets to circular orbits.
	
        The second simplification involves assuming that the orbital frequency of VBs remains static without any evolution. In reality, as GWDBs continuously emit GWs, their frequencies naturally escalate. Therefore, it becomes essential to explore how these fluctuations in frequency might impact subsequent research outcomes. Typically, LISA missions operate for a span of 4-5 years. Our main focus lies in evaluating whether the gravitational wave frequencies will exhibit substantial alterations within this 5-year time frame.
	
        We use \texttt{evol} in LEGWORK \citep{al} to calculate the SNR. Take source J0533+0209 as an example, its orbital period change in 5 years is $4.76\times10^{-4}\,\mathrm{s}$, the cyclical rate of change is $3.02\times10^{-12}\,\mathrm{s}/\mathrm{s}$, which is close to the results measured by \cite{ah}, i.e., $3.94\times10^{-12}\,\mathrm{s}/\mathrm{s}$. The ratio of the period variation to the orbital period is $3.86\times10^{-7}$, which can be considered as a constant period during the mission time period. In the subsequent processing, we will phase-fold the simulated data.
	
        We employ LEGWORK to simulate the GW signals. For a source, the total length of probes in one mission is in the range of 4-5 years, and we therefore simulate $ 1.2\times10^{8}\,\mathrm{s} $ time-scales. Since the orbital period scale is around $ 10^{3}\,\mathrm{s} $, to reduce the amount of data, we take the sampling frequency as $ 10^{-1}\,\mathrm{Hz} $, i.e. $ 10\,\mathrm{s} $ to measure one data point, in order to clearly visualize the GW signal.
	
        Then we add simulated GW noise onto the signals. We consider only the noise within a frequency window around the GW signal. The width of the frequency window $df$ is the frequency resolution $1/T$, where $T$ is the duration of the data. Here, we assume that the distribution of the noise is normal. Based on the derivation in the Appendix A, the standard deviation of the noise can be given as:
	
        \begin{equation}
		\sigma_{n}=\sqrt{S(f)\cdot\Delta f}.
	\end{equation}
	
        The sensitivity curve, as presented by \cite{am} and showcased in Figure \ref{sen curve}, delineates the correlation between noise strain and its respective frequencies. 
	
\begin{figure} 
            \centering
		\includegraphics[width=\linewidth]{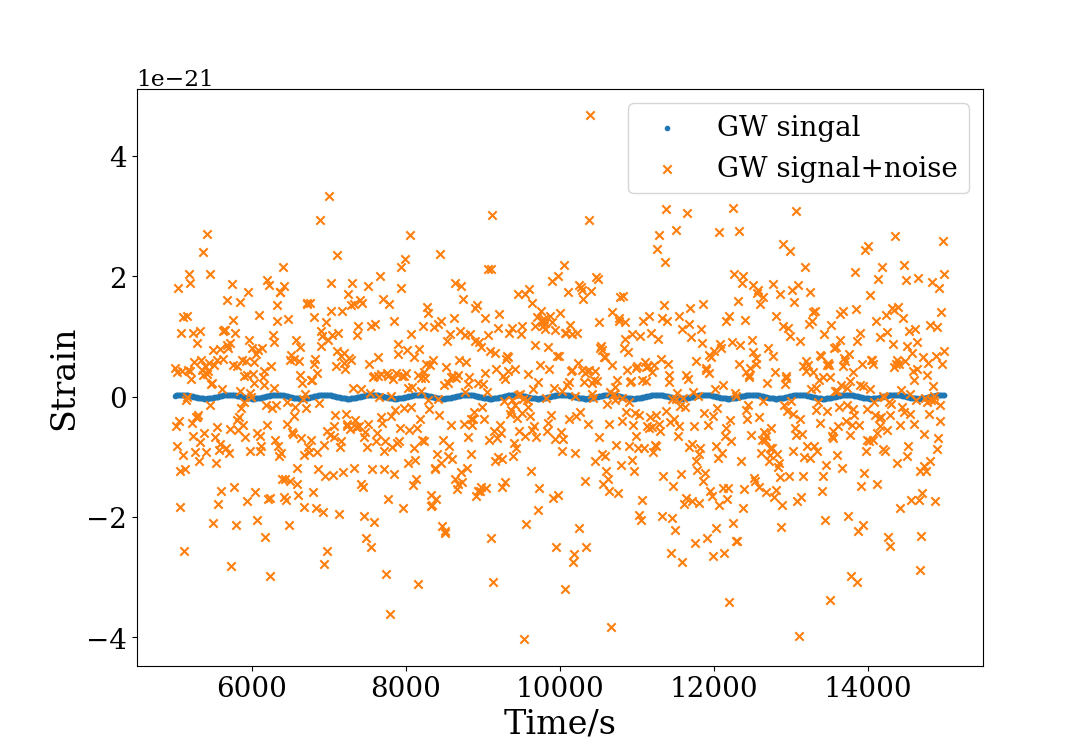}
            \caption{\textbf{The simulated GW signals within 5000-15000 seconds for J0533-0209 with SNR=15.2 in 4 years.}}
		\label{GW singal}
        \end{figure}

        In figure \ref{GW singal}, we plotted simulated signal and noise from 5000s to 15000s. The noise is filtered in the frequency range from $f_{\rm{GW}}-df$ to $f_{\rm{GW}}+df$, where $df=1/T$ is the frequency resolution of the data. In figure \ref{GW singal}, $df\sim10^{-4}\,\mathrm{Hz}$, while for the data we used below, $df\sim10^{-7}\,\mathrm{Hz}$. The corresponding SNR for this data segment is 15.2 in 4 years.

    \section{Simulation of optical signals} 
	
        For subsequent calculations, we require the optical data to vary sinusoidally as the GW, which excludes a fraction of the eclipsing binaries (though our method can be generalized easily to eclipsing binaries). A study conducted by \cite{w} in 2018 assembled a collection of 16 VBs. After eliminating specific eclipsing binaries and incorporating newly detected GWDBs based on the criteria previously discussed, the dataset was ultimately refined to consist of 14 VBs, as outlined in Table \ref{base data}.
	
        We also provided the basic information of the optical data of the VBs as shown in Table \ref{optical data}. To compare optical data quantitatively, we use relative root mean square error (RRMSE) \citep{as} to measure the quality of optical data:
	
        \begin{equation}
		R=\frac{\Delta A_{\mathrm{EM}}}{A_{\mathrm{EM}}\cdot\sqrt{N_{\mathrm{EM}}}},
        \end{equation}

        \noindent where $R$ is RRMES, $ A_{\mathrm{EM}} $ is the optical signal amplitude, $ \Delta A_{\mathrm{EM}} $ is the uncertainty, and $ N_{\mathrm{EM}} $ is the amount of data. RRMES is the total relative uncertainties of the data, so it can be used to qualify the accuracy of the optical data.
	
        Here, we directly take the optical data as a sinusoidal signal and superimpose on it the error of a normally distributed random function whose standard deviation is the uncertainty of the data. For the temporal distribution, we set the data points uniformly scattered randomly within the GW observation duration. 
	
        \begin{table*}
            \caption{\textbf{The limits on $\Delta c/c$ from simulated data}: The second column is the relative velocity difference using the optical r.v. data and the next two columns are the upper and lower limits of the 90\% confidence interval; columns 5-7 are the same as 2-4, but using a.m. observation instead of r.v. as EM data.}
            \small
            \renewcommand{\tabcolsep}{0.49cm}
            \renewcommand{\arraystretch}{1.5}
            \centering
            \begin{tabular}{ccccccc}
            \hline
            Source & $ \Delta c/c $ & lower limit & Upper limit & $ \Delta c/c $ & lower limit & Upper limit \\ 
            & (r.v.) & (r.v.) & (r.v.) & (a.m.) & (a.m.) & (a.m.) \\
            \hline
            J1539+5027 & $1.3\times10^{-11}$ & $-0.48\times10^{-11}$ & $3.0\times10^{-11}$ &   &   &  \\
            J0651+2844 & $0.27\times10^{-11}$ & $-4.7\times10^{-11}$ & $5.3\times10^{-11}$ &   &   &  \\
            J0935+4411 & $0.43\times10^{-10}$  & $-0.82\times10^{-10}$ & $1.7\times10^{-10}$ &   &   &  \\
            J0923-1218 & $-0.43\times10^{-10}$ & $-3.1\times10^{-10}$ & $2.2\times10^{-10}$ &   &   &  \\
            J1638+3500 & $0.27\times10^{-9}$ & $-2.3\times10^{-09}$ & $2.9\times10^{-09}$ &   &   &  \\
            J0130+5321 & $1.0\times10^{-10} $ & $-3.4\times10^{-10} $ & $ 5.4\times10^{-10} $ &   &   &  \\
            J1738+2927 & $0.52\times10^{-11}$  & $-1.9\times10^{-11}$ & $2.0\times10^{-11}$ &   &   &  \\
            J1115+0246 & $-0.21\times10^{-10}$ & $-4.0\times10^{-10}$ & $3.6\times10^{-10}$ &   &   &   \\
            J0533+0209 & $-0.16\times10^{-10}$ & $-1.3\times10^{-10}$ & $0.95\times10^{-10}$ & $0.060\times10^{-11}$ & $-3.5\times10^{-11}$ & $3.6\times10^{-11}$ \\
            J1401-0817 & $-0.16\times10^{-11}$ & $-1.2\times10^{-11}$ & $0.87\times10^{-11}$ &   &    & \\
            J1048-0000 & $-1.4\times10^{-11}$  & $-5.2\times10^{-11}$ & $2.3\times10^{-11}$ &   &  &  \\
            J0923+3028 & $-3.4\times10^{-10}$ & $-8.7\times10^{-10}$ & $1.8\times10^{-10}$ &   &   &  \\
            CD-3011223 & $-1.3\times10^{-10}$ & $-4.3\times10^{-10}$ & $1.8\times10^{-10}$ &   &   &  \\
            J1630+4233 & $1.4\times10^{-10}$  & $-0.34\times10^{-10} $ & $3.2\times10^{-10}$ &  &  &  \\ 
            \hline
            \end{tabular}
            \label{result}
        \end{table*}


    \section{Joint analysis and result}
	
\begin{figure} 
            \centering
            \includegraphics[width=.49\textwidth]{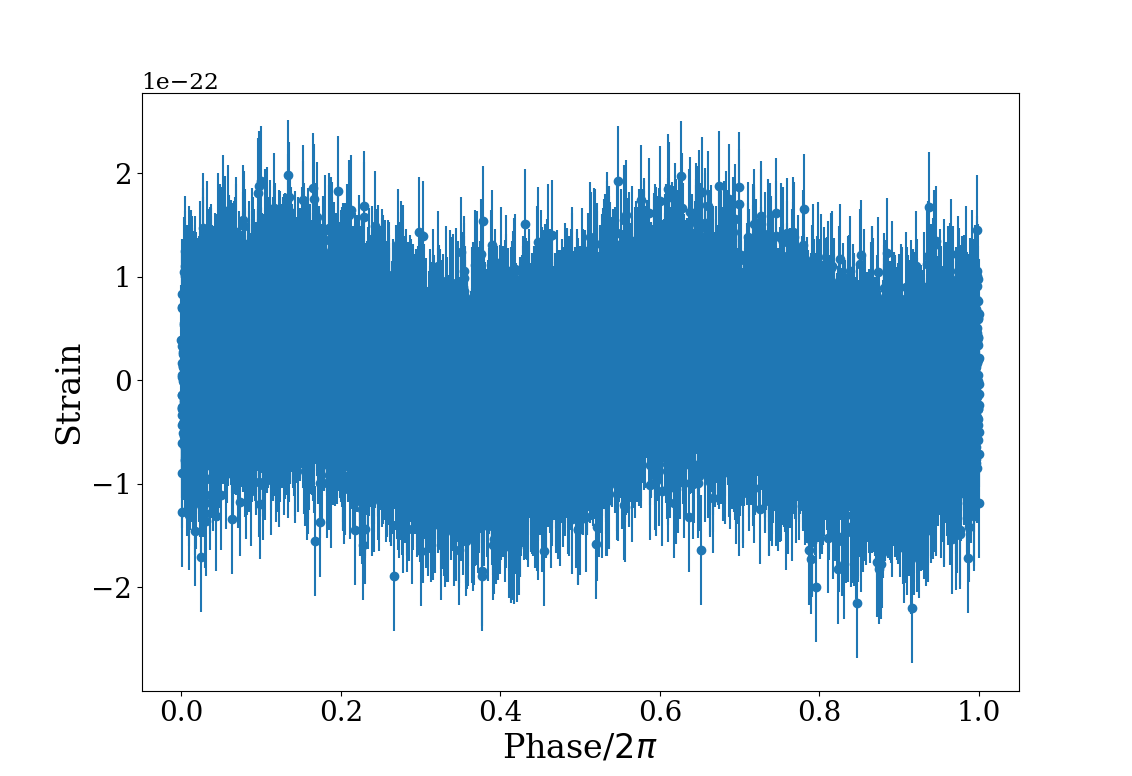}
            \includegraphics[width=.49\textwidth]{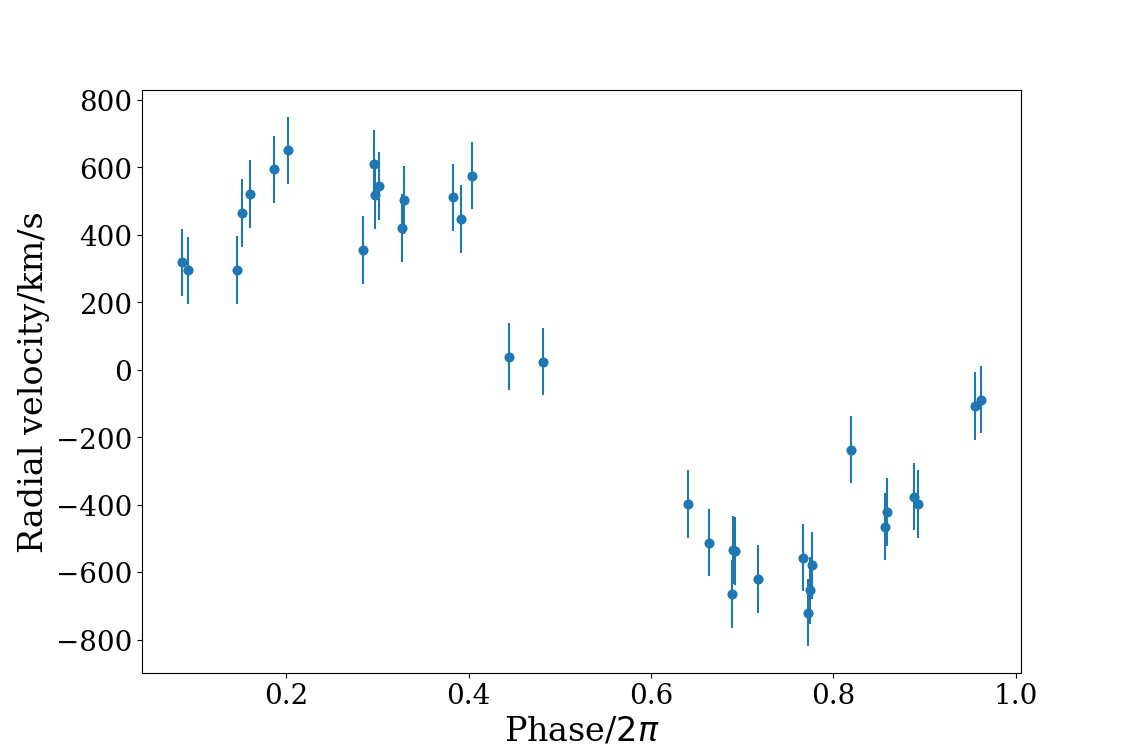}
            \includegraphics[width=.49\textwidth]{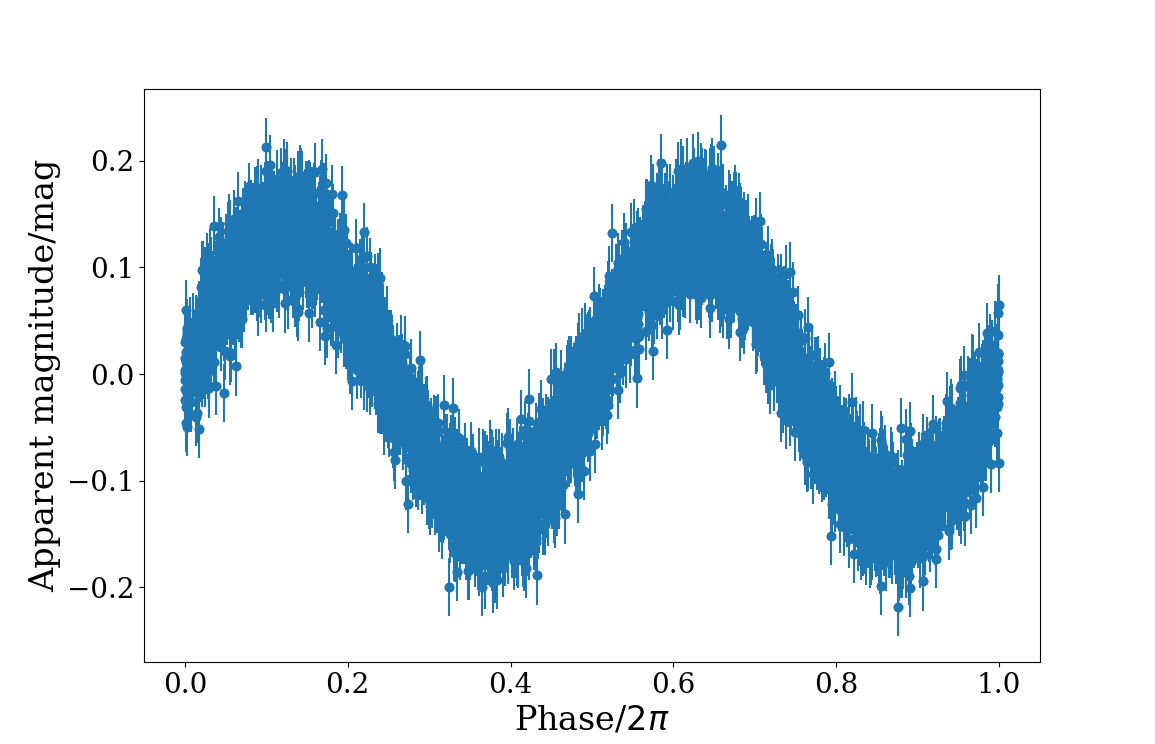}
            \caption{\textbf{Phase-folded signals with error bar}: Upper panel: GW data with $1.2\times 10^8$ s sampling frequency of $10^{-1}$ Hz; middle panel: 35 r.v. data uniformly scattered randomly within the GW observation duration; lower panel: 5000 a.m. data for the same distribution as r.v.. For clearer visualization, the GW panel has been down-sampled to 0.1\% of its original resolution. The r.v. measurement requires high quality spectrum and thus the data points are much fewer compared to that of a.m..}
		\label{phase}
	\end{figure}

        Due to the long time series of the simulation and the large time interval of the optical data, the direct fitting of the signal will produce greater errors. As discussed previously, both the GW data and the optical data have a sinusoidal waveform. Therefore, before formal processing, we first choose to phase fold the signal, i.e., use the obtained time series to invert the corresponding orbital phase value at that moment. The obtained data can demonstrate the measured quantities in one cycle, as shown in Figure \ref{phase}.
	
        Following this, we applied a curve fitting method to generate phase-shifted data along with the corresponding 90\% confidence interval. The fitting function utilized for this purpose is detailed as follows:
        
        \begin{equation}
		y = A\sin(n\cdot2\pi x+\phi),
        \end{equation}
        
        \noindent where $x$, $y$ are the phase and corresponding measurements, respectively, and $n$ is a constant introduced mainly due to the inconsistent periods of GW, r.v., and a.m. In particular, for GW and a.m. $n=2$ and for r.v. $n=1$. $A$ and $\phi$ are the values to be fitted, and the fitted phase $\phi$ is all we need.
	
        Here, we utilize fitting in order to obtain the expected value of the phase $ \mu $ with variance $\sigma^{2}$. Regarding the fitting process, the data distribution is ideally modeled by a T-distribution. However, considering that our dataset sizes are substantial, we approximate the following aspects using a normal distribution, as detailed in \citep{as}.
	
        For every source, we end up with at least two sets of data, i.e., the GW signal phase distribution $ \phi_{\mathrm{GW}}\sim N(\mu_{\mathrm{GW}},\sigma_{\mathrm{GW}}^{2}) $ and the optical signal phase distribution $ \phi_{\mathrm{EM}}\sim N(\mu_{\mathrm{EM}},\sigma_{\mathrm{EM}}^{2}) $. Therefore the distribution of the phase difference between the two should also be normal. Using Eq.\ref{final} we can derive the expressions for $\frac{\Delta c}{c}$ as:
        
        \begin{equation}
            \frac{\Delta c}{c}\sim N\big[\frac{Pc}{4\pi D}(\mu_{\mathrm{GW}} - n\cdot\mu_{\mathrm{EM}}),(\frac{Pc}{4\pi D})^{2}(n\cdot\sigma_{\mathrm{EM}}^{2} + \sigma_{\mathrm{GW}}^{2} + \sigma_{\varphi_{0}}^{2})\big].
        \end{equation}
	
        The 90\% confidence interval of the normal distribution is $ (\mu_{\Delta c/c}-1.645\sigma_{\Delta c/c},\mu_{\Delta c/c}+1.645\sigma_{\Delta c/c}) $. The upper and lower limits of the confidence intervals have been stated in Table \ref{result}.
	
        Based on the assumption that the noises of different VBs in different frequencies are independent \citep{at}, the joint probability of $\Delta c/c$ considering all these 14 sources is:
	
        \begin{equation}
            p(\frac{\Delta c}{c}|d_{1},d_{2}\dots d_{i})\propto p(\frac{\Delta c}{c}|d_{1})p(\frac{\Delta c}{c}|d_{2})\dots p(\frac{\Delta c}{c}|d_{i}),
        \end{equation}
	
        \noindent where $i$ is the number of independent constraints. Since there are 14 sources with both optical r.v. and a.m. data, the value of $i$ should be taken as $14$.
	
        For the product of multiple normal distributions, we determine the total expectation and variance using \citep{as}:
        
        \begin{equation}
		\begin{aligned}
                &\mu=\frac{\sum_{i}^{}\frac{\mu_{i}\prod_{m}^{}\sigma_{m}^{2}}{\sigma_{i}^{2}}}{\sum_{i}^{}\frac{\prod_{m}^{}\sigma_{m}^{2}}{\sigma_{i}^{2}}};\\
                &\sigma^{2}=\frac{\prod_{m}^{}\sigma_{m}^{2}}{\sum_{i}^{}\frac{\prod_{m}^{}\sigma_{m}^{2}}{\sigma_{i}^{2}}}.
            \end{aligned}
            \label{zhfc}
        \end{equation}

        Consequently, the final outcome is obtained as $-2.1\times10^{-12}\le\Delta c/c\le4.8\times10^{-12}$ ($-2.3\times10^{-12}\le\Delta c/c\le4.9\times10^{-12}$)\footnote{The results in the parentheses are those if we do not assume tidal synchronization of J0533+0209 and we use its r.v. instead of a.m.. We present the corresponding results in parentheses too in the next section.}.

    \section{Physics implication}
	
        Upon acquiring the constraints on $c_{\mathrm{GW}}$, we can establish limitations on the physical parameters as outlined in Section 1. The mass of the graviton can be obtained from Eq \ref{v-f} as:
	
        \begin{equation}
		m_{\mathrm{g}}=\frac{E}{c^{2}}\sqrt{-2\frac{\Delta c}{c}},\,\,\ c_{\rm{gw}}<c.
		\label{mass}
        \end{equation}
	
        The mass is related to the graviton energy $E=hf$. For GWDBs, the graviton energy is much lower than the GW events detected by LIGO due to their lower frequencies, which enables us to obtain more precise constraints. The distribution of graviton masses is derived by incorporating the constraints from each source into Eq. \ref{mass}. The best constraints involves utilizing source J1630-4233 with optical data of r.v. is $ m_{\mathrm{g}}\le3\times10^{-23}e\mathrm{V}/c^{2} $.
	
        Next, we delve into the examination of the coefficient $\bar{s}_{jm}$. If only the simplest case is considered, which means only $ \bar{s}_{00} $ is a non-zero coefficient, it holds: 

        \begin{equation}
		\bar{s}_{00}=\frac{-\frac{\Delta c}{c}}{\frac{1}{2}Y_{00}}.
        \end{equation}

        \noindent Substituting the results of the joint analysis yields $\bar{s}_{00}$ to $ -3.4\times10^{-11}\le\bar{s}_{00}\le1.5\times10^{-11} $ ($ -3.5\times10^{-11}\le\bar{s}_{00}\le1.6\times10^{-11} $), which is better than the result calculated in \cite{k}. For the general case, employing 14 values for substitution into equation \ref{SME} provides limitations for all $\bar{s}_{jm}$ coefficients. However, this particular calculation will be conducted in a distinct study.

    \section{Discussion}
	
        \begin{figure} 
        \centering
            \includegraphics[width=.49\textwidth]{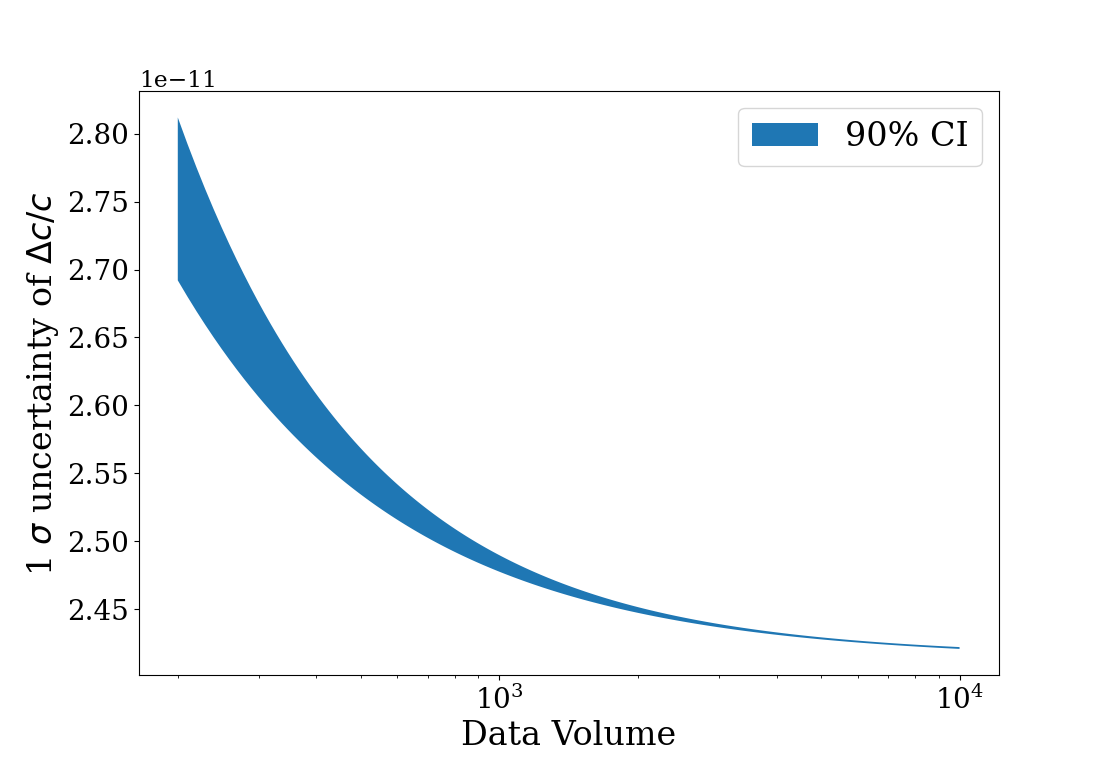}
            \includegraphics[width=.49\textwidth]{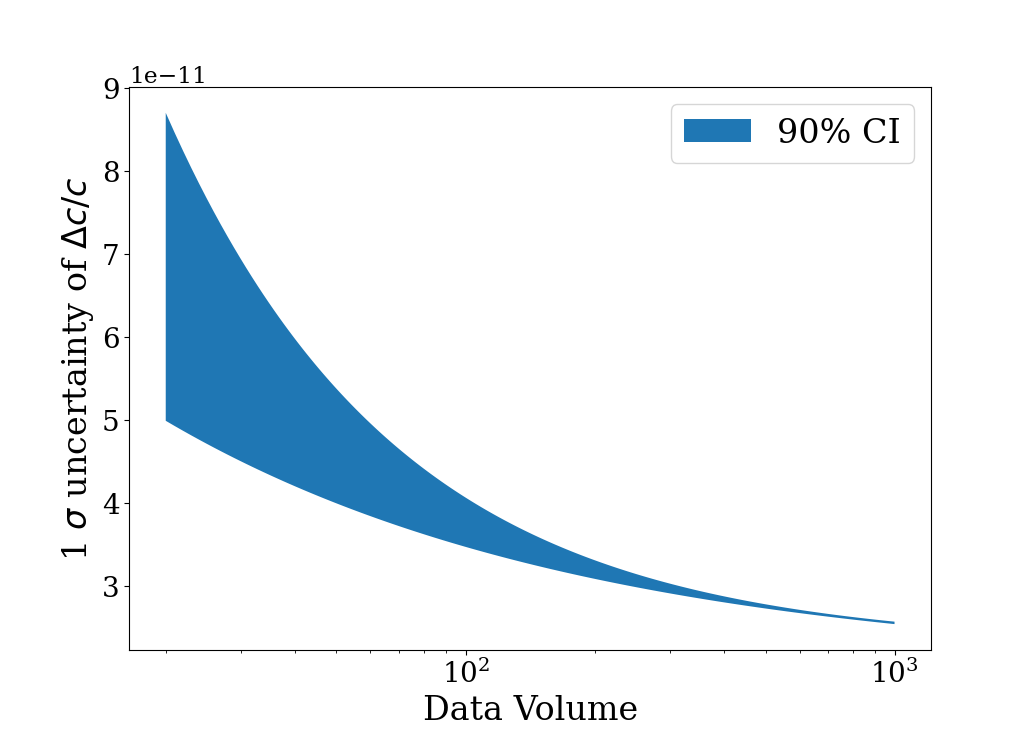}
            \caption{\textbf{Diagram of the uncertainty of $\Delta c/c$ after the change of optical data volume.}}
            \label{size}
            \end{figure}

        The methodology outlined in this paper showcases the potential to confine $\Delta c/c$ within the realm of $10^{-12}$, surpassing outcomes attained through GW dispersion. However, it falls short by approximately three to four orders of magnitude compared to the constraint utilizing the time lag between GW and GRB. In the subsequent discussion, we discuss different aspects that would strengthen this method.
	
\begin{figure} 
            \centering
            \includegraphics[width=.49\textwidth]{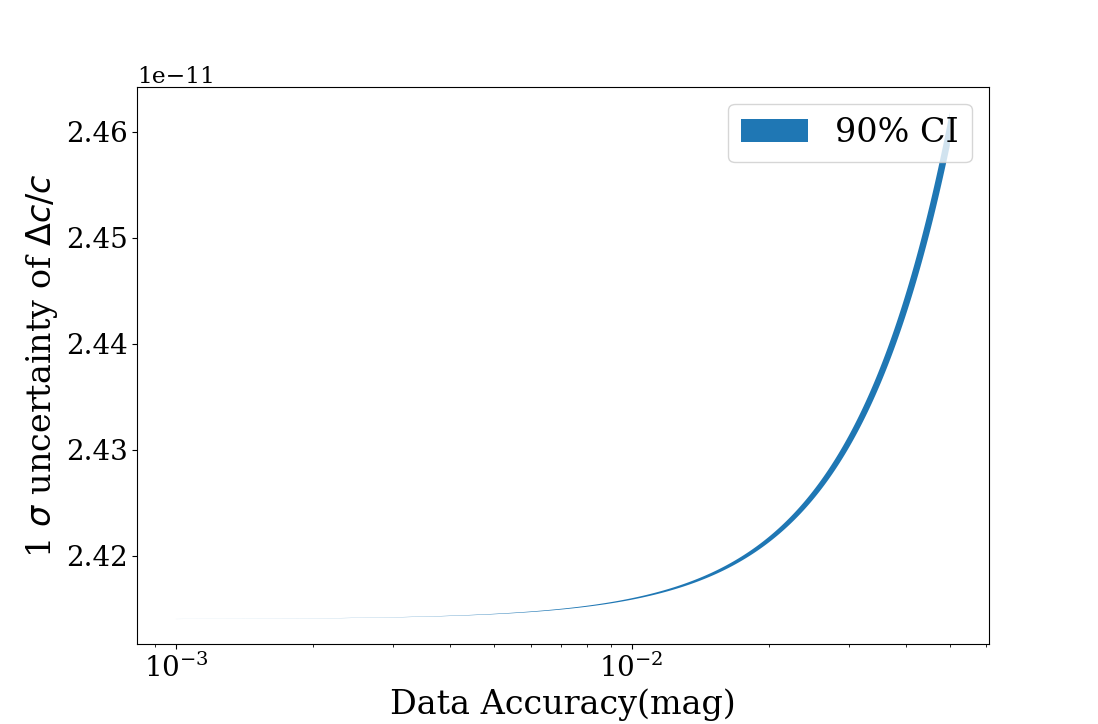}
            \includegraphics[width=.49\textwidth]{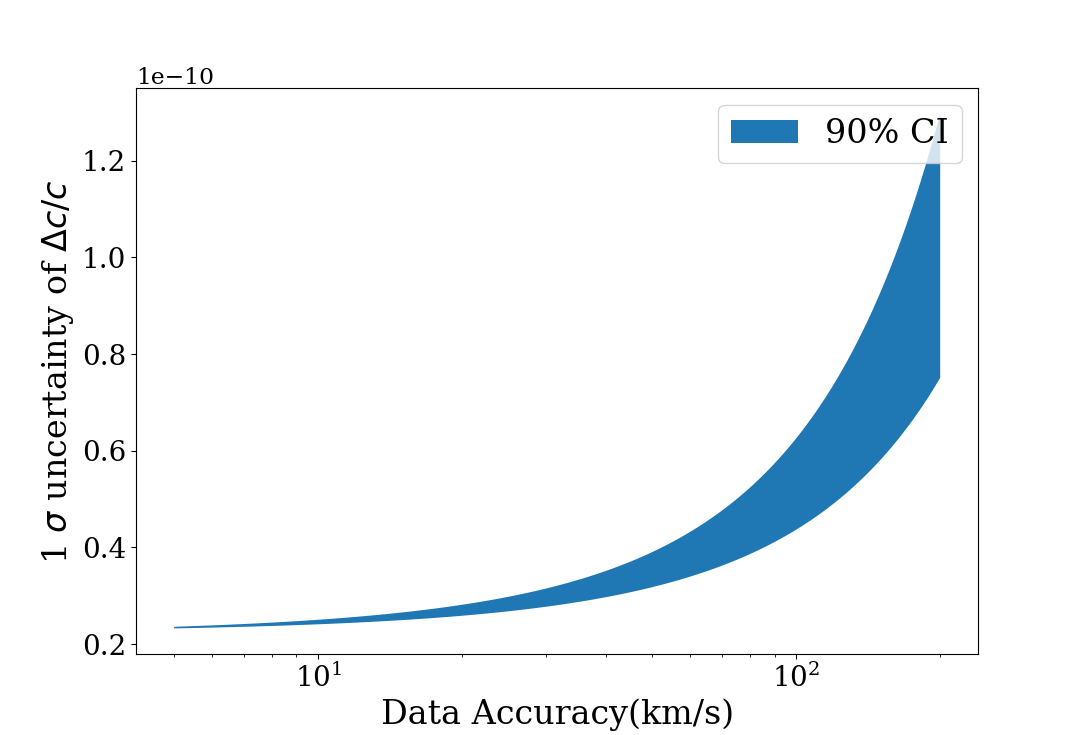}
            \caption{\textbf{Diagram of the uncertainty of $\Delta c/c$ after the change of optical data uncertainty.}}
            \label{uncertainty}
        \end{figure}

        As depicted in Table \ref{optical data}, the data sizes for r.v. typically tend to fall within the range of approximately 10 to 100, whereas the data sizes for a.m. typically scale up to hundreds or even surpass a thousand. Employing source J0533+0209 once more, we experimented by adjusting the data size for a.m. within the range of 100 to 10,000, while varying the data size for r.v. from 10 to 1,000. Employing the identical methodology as in the preceding section, each dataset was simulated 50 times to derive the distribution reflecting the ultimate standard deviation of the outcomes, as depicted in Figure \ref{size}. Similarly, we varied the uncertainties of r.v. and a.m. within two magnitudes and simulated each data set 50 times to obtain the distribution of the standard deviation of $\Delta c/c$ as shown in Figure \ref{uncertainty}. As the volume of data grows and quality enhancements are implemented, discernible improvements are evident in the simulation results, characterized by a decreased standard deviation and enhanced stability. Based on the above discussion, it is apparent that enhancing the quality of optical data may not significantly optimize the final results.

        Increasing the number of sources presents another viable method. With reference to the relationship described in equation \ref{zhfc}, we can roughly estimate the association between the number of sources, denoted by $ N_{\mathrm{s}} $, and the standard deviation of $\Delta c/c$ as:
 
        \begin{equation}
            \sigma_{\Delta c/c}\propto1/\sqrt{N_{\mathrm{s}}}.
        \end{equation}
	
        \noindent As LISA runs, more VBs will be discovered. In addition, the eclipsing binaries are not considered in this paper in order to simplify the model. In fact, if the theoretical function of the a.m. variation of the eclipsing binaries is known which can be accomplished using the \texttt{Phoebe} package \citep{bg}, it can also be fitted to constrain the phase difference. Based on our assessment, it seems plausible to include data gathered from at least 8 sources in the dataset.
	
        Overall, our method is astrophysical model independent which constrains $c_{\mathrm{GW}}$ with better accuracy. In addition, the use of more sources allows a more precise constraint on the parameter of local Lorentz violation. When analyzing the graviton mass, utilizing gravitational waves with lower frequencies leads to improved constraints on the mass.
	
        However, the final accuracy of this method is limited by the distance of the source. Compared to GRBs, whose distances are on the order of 10 Mpc, and VBs, whose distances are generally on the order of kpc, lead to final results that are 3-4 orders of magnitude larger than GRB. Also, a drawback of this one method is that the distance of the VBs cannot reach the Mpc magnitude, so the results obtained by optimizing the data are expected to be improved by only 1-2 orders of magnitude and cannot reach the accuracy obtained by GRB. 
	
        It is worth noting that $\phi_s=10^{-3}$ rad is the best case, while the average value of $\phi_s$ uncertainties is 0.1 rad. However, these uncertainties are estimated solely with GW data, while the sky position can be determined much better with optical observations, which is highly correlated with $\phi_s$ in GW inference. Therefore, the value reported from \cite{wa} should be considered as a conservative upper limit on $\delta\phi_s$.
	
        The method we proposed here can be easily transplanted into super black hole binaries (SBHB), where the distance can be in the order of 100Mpc, and thus results in constraint in $\Delta c/c$ orders of several orders of magnitude better. SBHB are believed to exist in the center of merged Galaxies, with masses from $10^{6}$ to $10^{10}$ solar masses. When the binary separation is close to pc scale, and the orbital period is $\sim$years, such systems are expected to emit GW in the nano-Hz band, which enters the range of Pulsar Timing Arrays (PTA) \citep{av}. On the other hand, SBHB also may exhibit themselves in EMW bands, as dual AGNs, or AGNs which show periodic light curves. There are two such examples: Mrk 915 and MCG+11-11-032 \citep{au}. They are believed to possess SBHB with masses $M\sim10^{8}M_{\odot}$ and $M\sim5\times10^{8}M_{\odot}$, red-shift $z=0.024$ and $z=0.0362$, distance $d=102.8\,\mathrm{Mpc}$ and $d=155.0\,\mathrm{Mpc}$ (Take the Hubble constant as $H_{0}=70\,\mathrm{km\cdot s^{-1}\cdot Mpc^{-1}}$), orbital period $T=35\,\mathrm{months}$ and $T=25\,\mathrm{months}$. The expected GW strain from them is $1.5\times10^{-16}$ and $1.9\times10^{-15}$. Which could be detectable with PTA in the near future \citep{aw}\footnote{We calculate the SNR with the software package \texttt{GW-Universe Toolbox \citep{ax}.}}.

\section*{Acknowledgements}

We would like to thank Prof. Yi-Ming Hu for his helpful comments and manuscript revision. This work is supported by the Chinese Academy of Sciences (Grant No. E329A3M1 and E3545KU2).

\section*{Data Availability}

The data underlying this article will be shared on reasonable request to the corresponding author.



\bibliographystyle{mnras}
\bibliography{main.bib}




\appendix

\section{\label{key}simulating the noise}
	
	For an arbitrary noise, the power is defined as:
	\begin{equation}
		P=\lim_{T\to\infty}\frac{1}{T}\int_{-\infty}^{\infty}|n(t)|^{2}\mathrm{d}t.
	\end{equation}
	Parseval's theorem tells us that we can re-write the average power as follows \citep{az}:
	\begin{equation}
		P=\lim_{T\to\infty}\frac{1}{T}\int_{-\infty}^{\infty}|n(f)|^{2}\mathrm{d}f.
	\end{equation}
	Then the power spectral density is simply defined as the integrand above \citep{ba}:
	\begin{equation}
		S(f)=\lim_{T\to\infty}\frac{1}{T}|n(f)|^{2}.
	\end{equation}
	Which means:
	\begin{equation}
		n(f)=\sqrt{S(f)\cdot T}e^{i\theta_{f}},
	\end{equation}
	where $\theta_{f}$ is a random phase \citep{ay}. Then the noise can be obtained using the inverse Fourier transform as follows:
	\begin{equation}
		n(t)=\mathcal{F}^{-1}[n(f)].
	\end{equation}
	
	However, if only consider noise within a small frequency $\Delta f$, we have:
	\begin{equation}
		\label{P(f)}
		\begin{aligned}
			P&=\int_{-\infty}^{\infty}S(f)\mathrm{d}f\\
			&=S(f)\Delta f
		\end{aligned}.
	\end{equation}
	Assume that the noise is normally distributed, we have:
	\begin{equation}
		\label{P(t)}
		\begin{aligned}
			P&=\lim_{T\to\infty}\frac{1}{T}\int_{-\infty}^{\infty}|n(t)|^{2}\mathrm{d}t\\
			&=\lim_{T\to\infty}\frac{1}{T}\int_{-\infty}^{\infty}E\big(|n(t)|^{2}\big)\mathrm{d}t\\
			&=\sigma_{n}^{2}\lim_{T\to\infty}\frac{1}{T}\int_{-\infty}^{\infty}\mathrm{d}t\\
			&=\sigma_{n}^{2}
		\end{aligned}.
	\end{equation}
	Combining Eq. \ref{P(f)} and Eq. \ref{P(t)}, we have:
	\begin{equation}
		\sigma_{n}=\sqrt{S(f)\Delta f}.
	\end{equation}
	
	The standard deviations of the noise simulated by the two methods are $5.3\times10^{-23}$ and $3.76\times10^{-23}$. Therefore we think the second method is simpler and accurate enough.


\bsp	
\label{lastpage}
\end{document}